 \documentclass[preprint2]{aastex}
 \usepackage{spr-astr-addons} 

\usepackage[usenames,dvipsnames]{color}
\usepackage{multirow}
\usepackage{subfigure}
\usepackage{subfig}

\begin{document}

\title{Stereo pairs in Astrophysics}

\shorttitle{Stereo pairs in Astrophysics}
\shortauthors{Vogt and Wagner}

\author{Fr\'ed\'eric Vogt\altaffilmark{1} and Alexander Y. Wagner\altaffilmark{1}}

\altaffiltext{1}{Mount Stromlo and Siding Spring Observatories, Research School of Astronomy and Astrophysics, Australian National University, Cotter Road, Weston Creek, ACT 2611, Australia.}

\begin{abstract}

Stereoscopic visualization is seldom used in Astrophysical publications and presentations compared to other scientific fields, e.g., Biochemistry, where it has been recognized as a valuable tool for decades. We put forth the view that stereo pairs can be a useful tool for the Astrophysics community in communicating a truer representation of astrophysical data. Here, we review the main theoretical aspects of stereoscopy, and present a tutorial to easily create stereo pairs using \textsc{Python}. We then describe how stereo pairs provide a way to incorporate 3D data in 2D publications of standard journals. We illustrate the use of stereo pairs with one conceptual and two Astrophysical science examples: an integral field spectroscopy study of a supernova remnant, and numerical simulations of a relativistic AGN jet. We also use these examples to make the case that stereo pairs are not merely an ostentatious way to present data, but an enhancement in the communication of scientific results in publications because they provide the reader with a realistic view of multi-dimensional data, be it of observational or theoretical nature. In recognition of the ongoing 3D expansion in the commercial sector, we advocate an increased use of stereo pairs in Astrophysics publications and presentations as a first step towards new interactive and multi-dimensional publication methods.

\end{abstract}

\keywords{Data Analysis and Techniques -- Tutorial; Stereoscopy -- Stereo pairs}

\section{Introduction}\label{Sec:intro}
\emph{Stereoscopy} consists in giving a depth perception out of 2D material to the viewer, and the concept behind it is fairly simple: it requires sending distinct and carefully chosen images to each eye, without one eye noticing the images intended for the other. The notion of depth perception, or \emph{stereopsis}, has been discussed as early as A.D. 280 by Euclid \citep{Okoshi76}. While there has been some experimentation using sketching techniques before 1800 \citep{Norling53}, the invention of photography by Niepce \citep{Smith1877,Perrier34} in the beginning of the 19$^{\mathrm{th}}$ Century marked the real start of extensive experimentation with stereoscopy. \cite{Wheatstone1838} assembled one of the earliest known stereoscopes, but Brewster is usually attributed the construction of the first practical viewing device, now referred to as the \emph{Brewster stereoscope} \citep{Norling53}.

As predicted by \cite{Scripture1899}, stereoscopy encountered quite a strong success in these early times, when stereoscopes where made widely available, because the production of stereo pairs become easier as photographic techniques evolved. \cite{Darrah77} discusses these early ages (from 1851 to 1935) of stereoscopy in depth, and we refer the interested reader to his work for a detailed overview of the various applications of stereo pairs in those times. 

Although the principle behind stereoscopy has remained the same ever since, there have been regular improvements to the methods of production and  visualization. In fact, the interest in stereoscopy has been closely linked to the development of both imaging and  visualization techniques \citep{Okoshi76}, and peaks of interest arose as new production and/or  visualization tools were invented. Beside the evolution of photographic techniques, the advent of computers and their ability to produce accurate and detailed stereo pairs represents one such development which resulted in a peak of interest for stereoscopy that started in the 70'. 

Several viewing devices have also been developed over the years, with one common aim: increasing the comfort and simplicity of stereoscopy for the viewer. In comparison to individual stereographs, the development of specialized glasses (red-blue, polarised, shutter-type) made stereo pairs easier to visualise. Lately, new stereoscopic technologies are being integrated into consumer products at a fast pace; movies, televisions, gaming consoles, cell phones, advertisement panels, and so on. Stereoscopy has become especially popular in the movie industry in the past few years with the advent of digital 3D cinemas. One should nonetheless not forget that stereo movies themselves are not recent: \emph{The Power of Love}, in anaglyph 3D, first aired in 1922 \citep[]{Zone07}.  

In the scientific community, stereoscopy has been known and used in the past, but the extent to which it has been exploited in the presentation of data differs from field to field. In Astrophysics, stereoscopy has not been used extensively, despite the multi-dimensional nature of many data sets. Often, a data cube is sliced or projected in order to obtain 2D publishable pictures and graphs. The issue of displaying and publishing multi-dimensional data sets has been identified in the past, and some interesting (non-stereoscopic) solutions have been proposed. Cosmologists working on the time evolution of the large scale structures in the Universe using 3D movies to illustrate their simulations results is one example \citep[e.g.][]{Holliman10}. Recently, \cite{Barnes08} described how documents in an \emph{Adobe Portable Document Format} (.pdf) are now able to contain animated 3D models, and described how this can be used to create interactive 3D graphs. In addition, \cite{Barnes06} developed a 3D plotting library specifically tailored to the needs of the Astrophysics community. \cite{Fluke06} also discuss and present alternative advanced image displays which might potentially take on a more significant place within the Astrophysics community in the future.

Yet, stereoscopic techniques are not unknown to Astrophysicists. Planetary scientists, for example, use red-blue anaglyphs. In the case of Mars \citep[e.g.][]{Neukum04,Keszthelyi08}, several probes and remote sensing satellites were equipped with special stereo cameras, for example the European Space Agency \emph{Mars Express} satellite and its High Resolution Stereo Camera Experiment \citep[][]{Jaumann07}, the \emph{Mars Reconnaissance Orbiter} and its High Resolution Imaging Science Experiment (HiRISE) camera \citep[][]{McEwen07}, the \emph{Phoenix} lander and its Surface Stereo Imager (SSI) \citep[][]{Lemmon08} or the imager for the Mars Pathfinder (MPI) mission \citep[][]{Smith97}.

Stereo pairs are another type of stereoscopic solution that has been employed to accommodate multi-dimensional data sets in publications. One of the first Astronomical stereo pairs published depicted the Moon and was created as early as 1862 by L.M. Rutherford \citep{Darrah77}. By taking two subsequent images of the Moon with a six days interval, he obtained a strong enough change in orientation to induce a reasonable feeling of depth. More recently, the advent of computers expanded the possible applications of stereo pairs in Astrophysics. For example, \cite{Yahil80} used them to display the position of the galaxies in the Revised Shapley-Ames Catalog; \cite{Martinet81} and \cite{Martinet81b} used stereo pairs to illustrate the shape of invariant surfaces in their study of dynamical problems with 3 degrees of freedom; \cite{Weygaert89} and \cite{Icke91} created stereo pairs to illustrate the Voronoi model they used to describe the asymptotic distribution of the cosmic mass on 10-200 Mpc scales; \cite{Rhee91} produced stereo pairs to show a 3D map of their sample of Abell clusters; \cite{Koutchmy92} published a stereo pair of the Solar Corona during the 1991 Solar eclipse; \cite{Sofue94} illustrated the stripping of the LMC H{I} and molecular clouds; \cite{Selman99,Selman99-3} created stereo pairs of the colour-magnitude diagrams of the ionizing cluster 30 Doradus; \cite{Sirko04} used stereo pairs to display the 3D position of the blue horizontal-branch (BHB) stars discovered in the SDSS \citep[][]{York00} spectroscopic survey; and \cite{Vogt10a} and \cite{Vogt10c} used stereo pairs in complement to their interactive 3D maps of the oxygen-rich material in SNR 1E 0102.2-7219 and SNR N132D. These examples do not represent an exhaustive list of all the work that has been published using stereo pairs in Astrophysics, but illustrate the wealth of topics that can profit and make use of this technique.
 
Nonetheless, the use of stereo pairs in Astrophysics is less prevalent than in other fields. In Biochemistry, for example, they have been an important tool to publish the 3D shapes of molecules from the beginning of the computer-era \citep[e.g.][]{Hamilton70,Hardman73} until today \citep[e.g][]{Pujadas01,Landsberg06, Xiang10}.

We believe that stereo pairs are a valuable tool in Astrophysics too, which recent 3D innovations may help renew. In other words, we argue that stereoscopy has a great but under-exploited potential for the publication of multi-dimensional Astrophysical data sets and can be a valuable complement to more standard plotting methods, especially with today's computing abilities. In Sec.~\ref{Sec:real}, after introducing the \emph{free-viewing} technique, we discuss the various theoretical ways to construct a stereo pair. In Sect.~\ref{Sec:tools}, we present our trade-off method to efficiently create stereo pairs of data cubes using \textsc{Python}. We then illustrate the unique features of stereo pairs as compared to standard plots in three different examples in Sect.~\ref{Sec:science}: a conceptual one in Sect.~\ref{Sec:conc}, one based on observational data in Sect.~\ref{Sec:N132D} and one based on theoretical data in Sect.~\ref{Sec:Alex}, for which the use of stereo pairs provides critical scientific information. We discuss the role that stereo pairs can play regarding future developments of 3D visualization techniques in Sec.~\ref{Sec:future}, and summarize our conclusions in Sect.~\ref{Sec:concl}.

\section{A theoretical overview of stereo pairs}\label{Sec:real}
Several terms, such as \emph{stereoscopy, stereopsis, stereo pairs} and \emph{stereograms}, are used throughout the literature, sometimes with slightly different meanings. To avoid confusion in this article, we refer to; \emph{stereopsis} as the depth perception reconstructed by the brain, \emph{stereoscopy} as the science of inducing depth perception using 2D material of any type, \emph{stereograms} as any type of image capable of inducing a depth feeling when viewed with the proper material, and \emph{stereo pair} as a specific type of stereogram, where the left hand side (the image for the left eye; LHS) and right hand side (the image for the right eye; RHS) images are located side-by-side. 

\subsection{Visualizing a stereo pair: the free-viewing technique}\label{Sec:free-viewing}
At any time, our brain interprets the simultaneous images from each of our eyes, combines them and automatically reconstructs a 3D image. The reconstruction algorithm is completely unconscious, and it feels natural to see our environment in 3D. If one looks at a structure at a distance of $\sim$60 cm for example, it will be seen by the right eye as if it had been rotated by $\sim$5 degrees as compared to the left eye's view. Hence, the very same 3D feeling can be achieved by sending two different 2D pictures of the same object to the left and right eye, if the pictures are taken from two different view angles. As long as the brain believes that it is looking at a \emph{real} object, its automatic 3D reconstruction algorithm will work. The main challenge is to be able to provide a different image to each eye without the other noticing it. 

As we mentioned previously in Sect.~\ref{Sec:intro}, several techniques exist to achieve this goal, such as stereoscopes, glasses, or auto-stereoscopic screens (for which no glasses are required). Depending on the visualization technique, stereoscopic images carry different names : anaglyphs (red-blue, polarized) need to be looked at using the appropriate glasses; stereo pairs, and autostereograms \citep{Tyler90} require the so-called free-viewing technique. This latter way of looking at stereo pairs has the advantage that it does not require any special equipment, provided that the stereo pair is in the correct format. This advantage makes it the best method to visualize stereo pairs in publications, as a majority of the readers accessing the article in its online or printed version will be able to directly get a feeling of depth. With the left and right images side-by-side, it is up to the reader to have each eye looking at one image only, thus recreating the 3D feeling. Obviously, this requires the reader to be familiar with the technique. However, the extended usage of stereo pairs in other fields, as mentioned in Sect.~\ref{Sec:intro}, gives us confidence that this fact should not represent an obstacle to the popularization of stereo pairs in Astrophysics. The web offers a vast resource of examples for training one's ability to see the 3D images from stereo pairs rapidly, and many websites provide tutorials and suggestions on how to make it work \footnote{for example \textsf{http://spdbv.vital-it.ch/TheMolecularLevel/0Help/StereoView.html\#con}}. 

\begin{figure}[htb!]
\centerline{\includegraphics[scale=0.5]{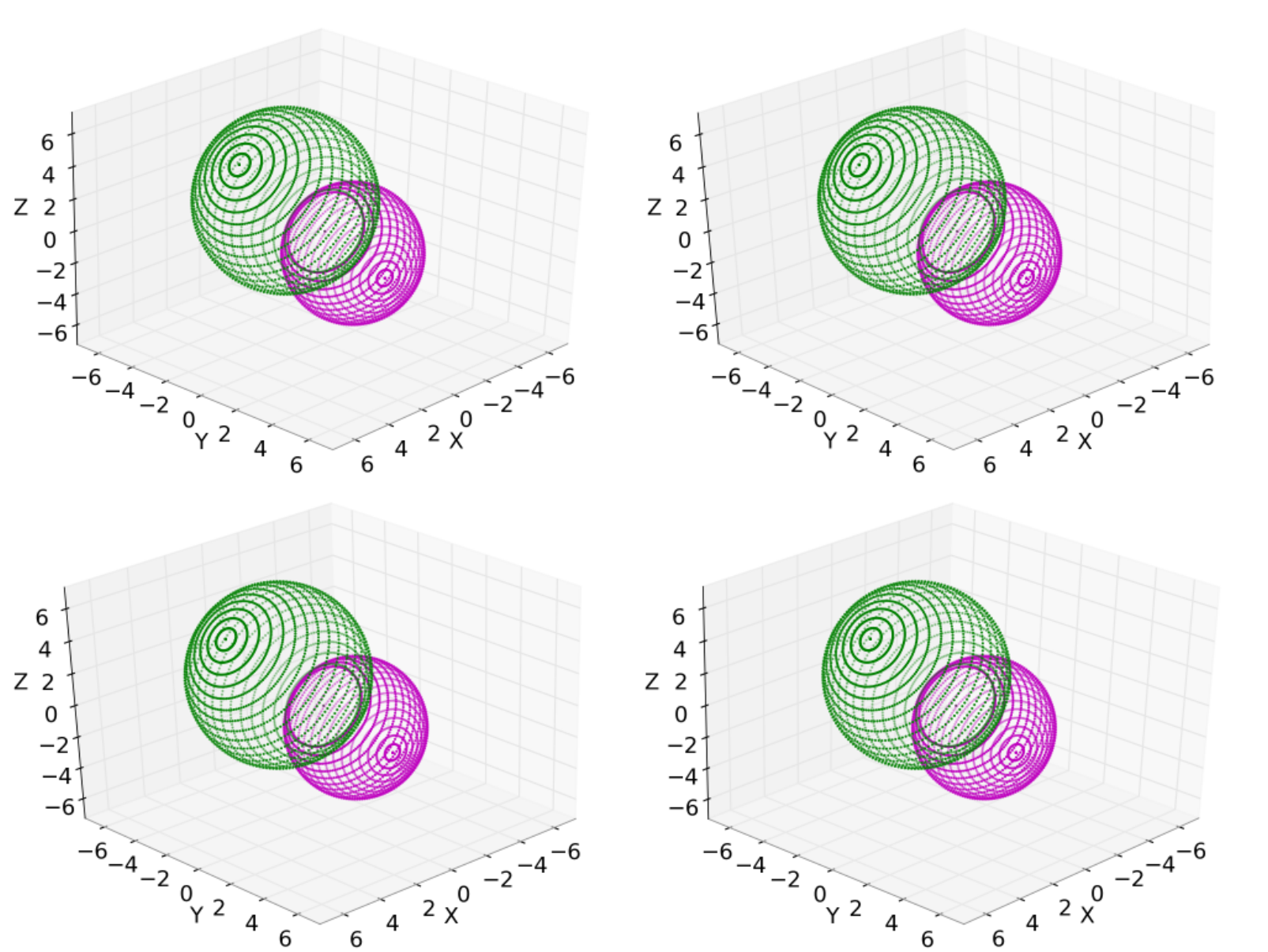}}
\caption{Parallel (top) and cross-eyed (bottom) stereo pairs of two intersecting spheres of different radii.}\label{Fig:para_cross_sphere}
\end{figure}

There are two ways to look at stereo pairs : parallel and cross-eyed. The names refer to the required orientation of the eyes, which depends on the position of the LHS and RHS images within the pair: parallel viewing requires the LHS image to be on the left (and the RHS image on the right) while cross-eyed requires them to be swapped. Whether one technique is more comfortable than the other is a matter of personal opinion and training. In Fig.~\ref{Fig:para_cross_sphere}, we present for comparison two stereo pairs of the same object - two intersecting spheres of different radius - from the same viewpoint. The top pair is designed for parallel viewing, while the bottom one requires the cross-eyed technique to be visualized properly.

Looking at a stereo pair for the first time can be quite challenging and there exists several ways to make it work. After experimenting with some students and astronomers of the Mount Stromlo Observatory, many having no previous experience with stereo pairs, we provide some suggestions that might help when looking at a stereo pair for the first time :
\begin{itemize}
\item[\textbf{Parallel viewing :}] Holding the printed page in front of you, look at the horizon. Once in focus, lift up the page so that the stereo pair reaches the level of your eyes, but without adjusting the focus onto the page - the focus remains at the horizon, however, your \emph{attention} is on the page. If done correctly, the left and right images will merge into a central 3D image of the double sphere. If your eyes focus on the page when you movie it up, try to relax them. If you find it difficult to see the 3D image (so that you can read the axis labels), it might help to convince yourself that you are \emph{looking at something real}.
\item[\textbf{Cross-eyed viewing :}] Place the printed page on a flat surface, and position your head straight above the stereo pair. Bring one finger in between the page and your eyes, so that its tip is located just below the pair, in the middle. Focusing on your finger tip will merge the background left and right images into a central 3D one. If they do not merge properly, try adjusting the height of your finger. The last step consist in removing the finger while keeping the central 3D image of the stereo pair, and might require some concentration/practice.
\end{itemize}

For both techniques, the alignment of the head is important - a slight rotation will have for consequence that the LHS and RHS will not overlap properly when merging them. With some experience, stereo pairs can be viewed on paper as well as on a computer screen. Our own experience as well as that of astronomers and students at the Mount Stromlo Observatory show that getting the eyes in the correct position becomes easier with practice, ultimately becoming almost an automatic adjustment.

In this article, we shall only use stereo pairs designed for parallel viewing, which is the most comfortable and natural technique for most of the people we have asked around us. Note that looking at a stereo pair with the wrong technique will have the consequence of reverting the depth axis (i.e. flipping the object back to front). While using the appropriate viewing technique is recommended, this fact nevertheless enables people more comfortable with the cross-eyed technique to obtain a 3D impression of every stereo pairs within this article. Generally, stereo pairs in Biochemistry publications are of the parallel type, a tradition which is most likely a remnant of the necessity to accommodate the use of viewing devices in the 70's, as described by \cite{Smith71}. There is however no fixed rules, and the stereo pair constructed by \cite{Sirko04} of the position of BHB stars in the SDSS spectroscopic survey is for example a cross-eyed pair.

\subsection{Constructing a stereo pair: the Toe-in and Offset methods}

\begin{figure}[htb!]
\centerline{\includegraphics[scale=0.3]{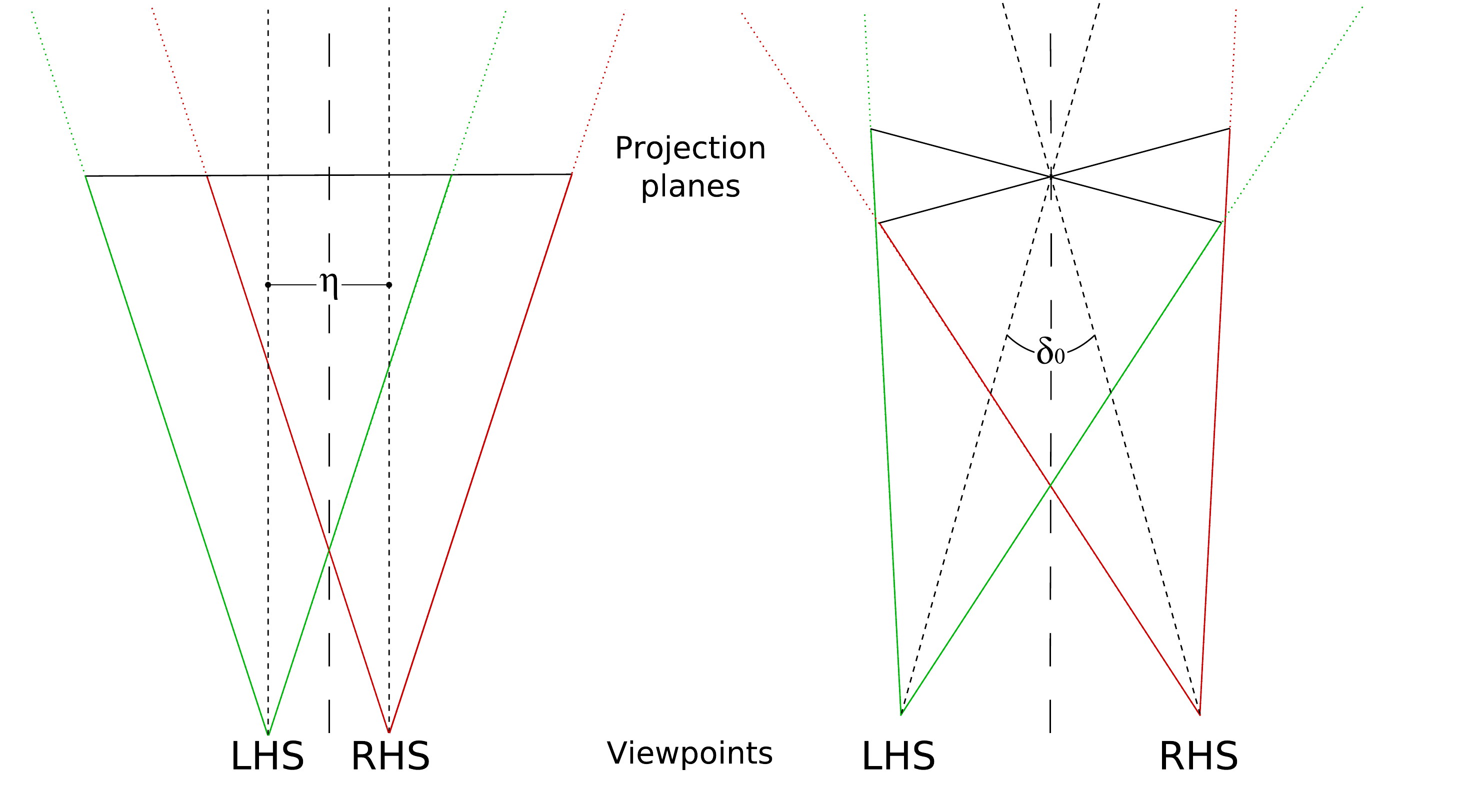}}
\caption{Geometry for the construction of Offset (left) and Toe-in (right) stereo pairs. LHS and RHS projections of the original 3D data set are created along two different view vectors.}\label{fig:view}
\end{figure}

There exist two main techniques to build a stereo pair, depending on the projection method used to obtain the left and right images. The first one, known as the \emph{Toe-in} method, mimics the human eyes' behaviour, and is in that sense the most obvious technique. The concept, illustrated in Fig.~\ref{fig:view}, is as follows : the LHS and RHS projections of the 3D scene are created by projecting it along two view vectors rotated by $\delta_0=\sim$2-5$^{\circ}$ along the azimuthal angle. The value of $\delta_0$ is somewhat arbitrary, and a higher angle will result in an increased depth perception for the viewer. Our own experimentations, as well as various online tutorials, indicate that a value of $\delta_0=$5$^{\circ}$ works well. This freedom in the choice of $\delta_0$ is also reflected in Biochemistry stereo pairs, for which several values are being used throughout the literature; 2$^{\circ}$ \citep[]{Hayman87}, 3$^{\circ}$ \citep{Berry10}, 5$^{\circ}$ \citep{Robinson89, Stockert94}, 6$^{\circ}$ \citep{Hayman87,Stockert94} and 10$^{\circ}$ \citep{Stockert94}.

There are several issues with the Toe-in method, one of the principal being vertical parallax due to keystoning. Because the projections are taken at an angle, successive planes perpendicular to the viewer will be slightly distorted. The distortion is inverted in the LHS and RHS images. As a result, when merging the two images, there will be a mismatch in the outer points, resulting in a blurred image. It should be noted that keystoning is closely related to the distance of the camera to the object, and is more important when being close from the 3D scene. The second issue with the Toe-in method lies in the fact that the eyes will have to adjust from a convergent to a divergent position in order to scan the depth dimension. For extended objects this can result in strong discomfort for the viewers, and even an impossibility to merge the left and right images if the converging/diverging angle becomes too important. \cite{Woods93} discuss the various distortions present in stereo pairs in great details, and we refer the interested reader to his article for more details.

The Offset (or Off-axis) method addresses and solves the problems of the Toe-in projection, and is in that sense sometimes considered to be more \emph{correct}. In this case, the LHS and RHS \emph{cameras} look at the 3D scene in parallel directions, and offset by a distance $\eta$. This method does not create any keystoning, and ensures that the eyes remain in the same position when scanning the depth dimension of the reconstructed 3D picture. An illustration of the method is shown in Fig.~\ref{fig:view}. One of the inherent drawbacks of this method is that the outside regions of the LHS and RHS fields do not overlap - and hence must be taken off the final image.

Choosing between the Toe-in and the Offset method to create stereo pairs is entirely up to the creator of the pair. As we will discuss in the next Section, the Toe-in method can provide excellent stereo pairs under certain circumstances ; the stereo pairs shown in Fig.~\ref{Fig:para_cross_sphere} have for example been created using the Toe-in projection method, and in the Biochemistry literature, many tutorials describing the creation of stereo pairs use the Toe-in method \citep[e.g.][]{Hayman87,Robinson89,Stockert94,Berry10}. Furthermore, sometimes, intrinsic limitations of the software or programming language used to create the stereo pair might require a trade-off between feasibility and quality, as we will show in Sec.~\ref{Sec:tools}.

The last step required to construct a stereo pair consists in placing the LHS and RHS images side-by-side. The distance between the two images is not critical. This reflect the fact that the interpupillary distance is not uniform across the population, but varies with age, gender, and race. Specifically, \cite{Dodgson04} mentions a mean interpupillary distance of 63~mm for adults, with the vast majority lying within a 50-75~mm range. In this article, we have used point-to-point separations (between the LHS and RHS images) ranging from 3.5~cm to 5~cm depending on the stereo pair, and it is a matter of personal opinion as to which value is most comfortable. Increasing the inter-image distance above 6~cm is not advisable, as the stereo pair might become harder to visualize for people with a smaller interpupillary distance than average.  

\section{Constructing a stereo pair : a \textsc{Python}/\textsc{Matplotlib} solution}\label{Sec:tools}

For stereo pairs to be recognized as a valuable tool by the Astrophysics community requires them to be extremely easy to create, implement and link to the data set. Let us consider the 3D data set used to produce Fig.~\ref{Fig:para_cross_sphere}, which can be seen as a cloud of points in 3D space. There exist many methods in order to easily produce stereo pairs from such a data cube, and it would be impossible to list them all here. Some commercial software packages can produce stereo pairs with a single mouse click - the stereo pairs shown in Sec.~\ref{Sec:Alex} were for example produced with the \emph{VisIt} software. Alternatively, many scientists have developed their own software, specifically tailored to their own data type and format. Such customized or commercial software are capable of creating excellent stereo pairs, and are usually designed well enough not to have too steep learning curves. 

Here, we propose an alternative, off-the-shelf, efficient way to produce stereo pairs using the programming language \textsc{Python}. This method, which we will refer to as the \emph{simplified Toe-in} (sTi) is based on our own experimentation, and is extremely straightforward to implement, even for people with little/no experience with this language. Furthermore, creating stereo pairs with \textsc{Python} grants access to a large collection of non-plotting modules to work on the data cube beforehand. This provides the more advanced user with the freedom to potentially reduce, sort, fit and clean the data set before creating a stereo pair - a strong advantage as compared to commercial software which often requires specific input, and does not enable direct data interaction.

Our sTi method is based on the Toe-in projection described previously, but accounts for current limitation in the \textsc{Matplotlib} plotting module, in which several projection parameters are currently hard-coded and cannot be accessed easily. The required functions are located within the \textsc{mplot3d} toolkit\footnote{see the online documentation for a detailed description : \textsf{http://matplotlib.sourceforge.net/mpl\_toolkits/mplot3d/index.html} }, and are designed to create a 2D projection out of a 3D scene. Specifically, the \textsc{Axes3D} instance, creating the plotting area, takes in two parameters, the elevation $\theta$ and the azimuth $\phi$, both in degrees. Creating an sTi stereo pair is then a 3-steps process :
\begin{enumerate}
\item[1:] Create two side-by-side (using \textsc{subplot}) plots, each with the \textsc{Axes3D} instance.
\item[2:] For a viewpoint located at ($\phi_0$;$\theta_0$), set the left plot viewpoint to ($\phi_0-2.5^{\circ}$;$\theta_0$) and the right plot viewpoint to ($\phi_0+2.5^{\circ}$;$\theta_0$)
\item[3:] Print the data using the appropriate \textsc{mplot3d} function, such as \textsc{scatter3D} for a cloud of points, or \textsc{contour3D} for isosurfaces.
\end{enumerate} 

This method is a trade-off : Toe-in or Offset stereo pairs cannot presently be created with \textsc{Matplotlib} easily, unlike sTi stereo pairs. Especially, implementing the Toe-in method requires the modification of the source code of the \textsc{Axes3d} function\footnote{We hope to eventually include our source code update in an upcoming release of \textsc{Matplotlib}. In the meantime, a copy of our modified \textsc{Axes3d} function, that enables the creation of Toe-in stereo pairs, can be obtained from \textsf{fvogt@mso.anu.edu.au}}.

But the sTi simplicity comes at a price. For viewpoints with no elevation ($\theta_0=0^{\circ}$), the sTi method is identical to the Toe-in technique. However, errors both in the LHS/RHS images orientation, as well as in their azimuthal separation, are introduced with increasing values of $|\theta_0|$. For completeness, we describe those issues in detail in the Appendix~\ref{App}, and compare sTi stereo pairs with Toe-in and Offset stereo pairs for different elevations. The comparisons show that the sTi method delivers very similar results to the Toe-in method for elevation as high as $|\theta_0|\sim50^{\circ}$. Beyond this limit, the depth perception is reduced compared to the Toe-in method. Comparing with the Offset method reveals that the 3D structure of the object is increased and better revealed in the sTi method, which makes the latter more suitable for the publication of multi-dimensional data sets. In other words, if the object appears, with the Offset technique, to be \emph{popping out} of the screen, it does not contain itself much depth information. \cite{Peterka09} reached a similar conclusion when building a stereoscopic movie of 3D simulations of a core-collapse supernova \citep[][]{Blondin03} : " \emph{ [The Offset technique] ... created a plausible facsimile of 3D. However, the trained observer noticed the flatness in the center of the sphere, and we did not want to rely on 2D depth cues such as lighting and shading to convey 3D information.}" Hence, the Toe-in, and in our case the sTi technique, is recommended for the creation of stereo pairs in Astrophysics. It is: 
\begin{itemize}
\item more efficient at providing a depth structure to the data as compared to the Offset method.
\item easier to implement in \textsc{Python} than the proper Toe-in method (which requires the modification of the source code).
\end{itemize}

The hard-coded parameters within \textsc{mplot3d} cause no visible vertical parallax or other visual defects in sTi stereo pairs.

We have asked some students and astronomers at the Mount Stromlo Observatory to test our sTi stereo pairs. None of them found the sTi pairs more tiring or difficult to visualize, with a very satisfactory depth impression. Most of them also noted that even if the depth perception in sTi pairs is degraded beyond $|\theta_0|\sim50^{\circ}$ compared to Toe-in pairs, it does not vanishes completely. In that sense, our suggested sTi method can be used for any viewpoint with satisfactory depth impression and reasonable comfort for the viewer.

\section{Application examples for stereo pairs}\label{Sec:science}
We illustrate the role that stereoscopy, in the form of stereo pairs, can play in a publication with three examples; conceptual, observational, and theoretical. Clearly, the range of applications for stereoscopy in Astrophysics is much larger than those we are about to present, and the publications mentioned in Sec.~\ref{Sec:intro} highlight the fact that stereo pairs can be used for almost any type of multi-dimensional data set, e.g., 3D maps and structures, 3D iso-surfaces, N-body simulations and trajectories, cosmological simulations, magnetic and other field maps, 3D function fitting, color-magnitude diagrams, hydrodynamic simulations and complex (e.g. turbulent) structures.

\subsection{Conceptual example: intersecting spheres}\label{Sec:conc}
\begin{figure}[htb!]
\centerline{\includegraphics[scale=0.4]{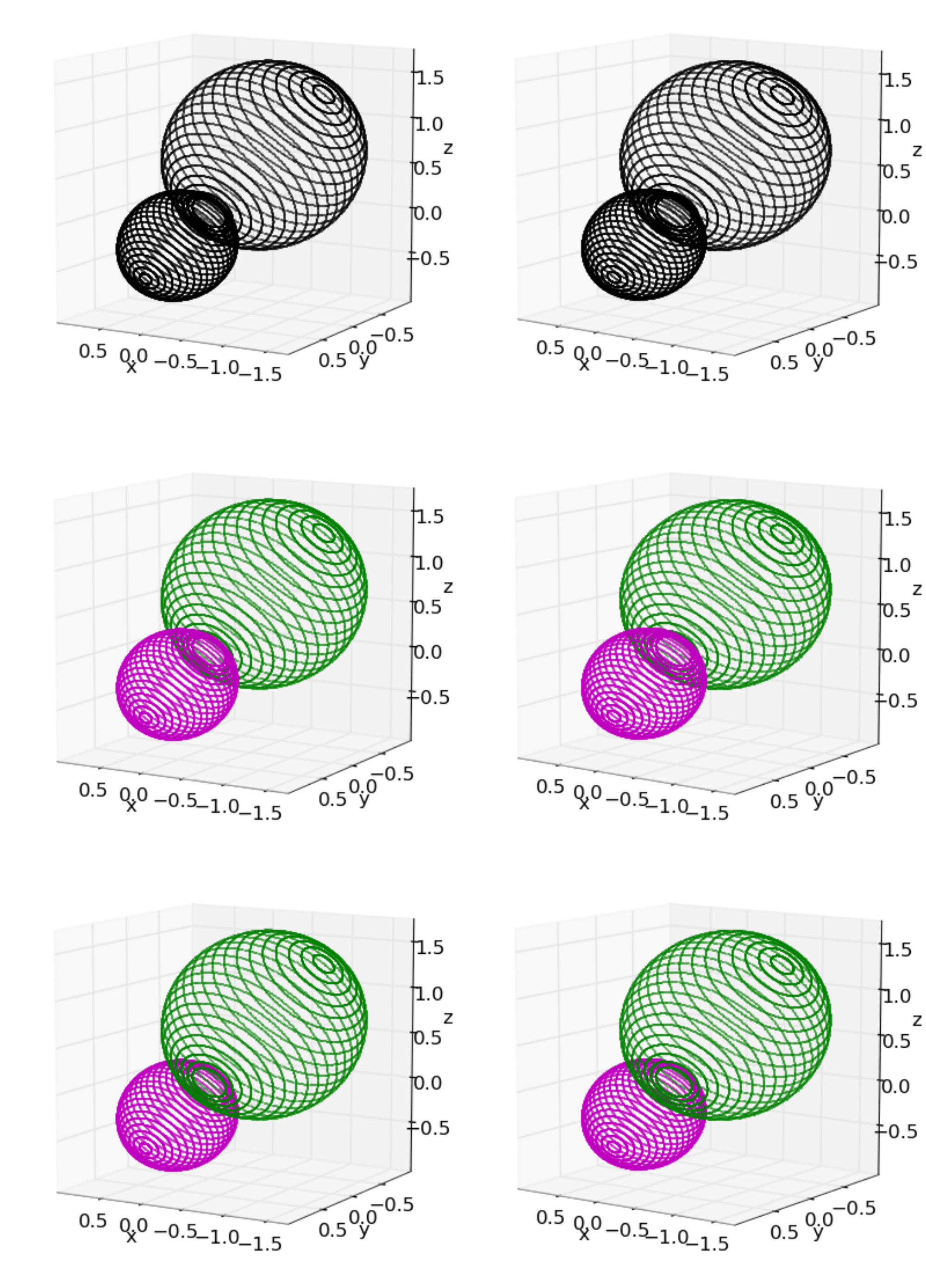}}
\caption{Stereo pairs of two intersecting spheres of different radius, shown in black and white (top) and color (middle and bottom). In each case, the two spheres orientation axis is in a XZ plane, and tilted at 45 degrees with respect to the Z axis. The orientation is best revealed in the 3D reconstruction using the free-viewing technique.}\label{fig:spheres}
\end{figure}

As mentioned in Section~\ref{Sec:intro}, stereoscopy, in this case in the form of stereo pairs, is different from standard plots in that it transmits a feeling of depth to the viewer. Let us illustrate this advantage with a practical example. In Fig.~\ref{fig:spheres}, we present three stereo pairs of the same object, two intersecting spheres of different radius. In each case, the two spheres' symmetry axis is in the XZ plane, and tilted by 45 degrees with respect to the Z axis. 

Several observations can be made at this point. First, stereo pairs work both in color or greyscale. Second, stereo pairs provide the reader with a true feeling of depth, and unambiguously convey the orientation of the structure. In the case of this double-sphere, the use of a stereo pair removes the ambiguity that arises when looking at a single image only, in which case the spheres could be seen as being oriented in the XZ or YZ plane. The use of colors may also help lift the orientation degeneracy: in the bottom pair, plotting the green sphere \emph{above} the pink one does suggest that the axis lies in the XZ plane. However, in some cases, it might not always be possible to define the order in which the objects are plotted (or printed) by color. In the middle pair, we have intentionally plotted the pink sphere above the green, which then suggest, when looking at only one of the image, that the rotational axis lies in a YZ plane. In that case, stereo viewing is perfect to correct this wrong feeling. In 3D, the middle pair looks essentially identical to the bottom one, with the rotational axis lying in the XZ plane (i.e. the big, green sphere is on top in all three cases). While this example is rather simple, it nevertheless illustrates some key advantages that stereo pairs have over single projected images. We reinforce this point with stereo pairs of more complex shapes presented in the next sections.

\subsection{Observational example : SNR N132D in the LMC}\label{Sec:N132D}

\begin{figure}[htb!]
\centerline{\includegraphics[scale=0.55]{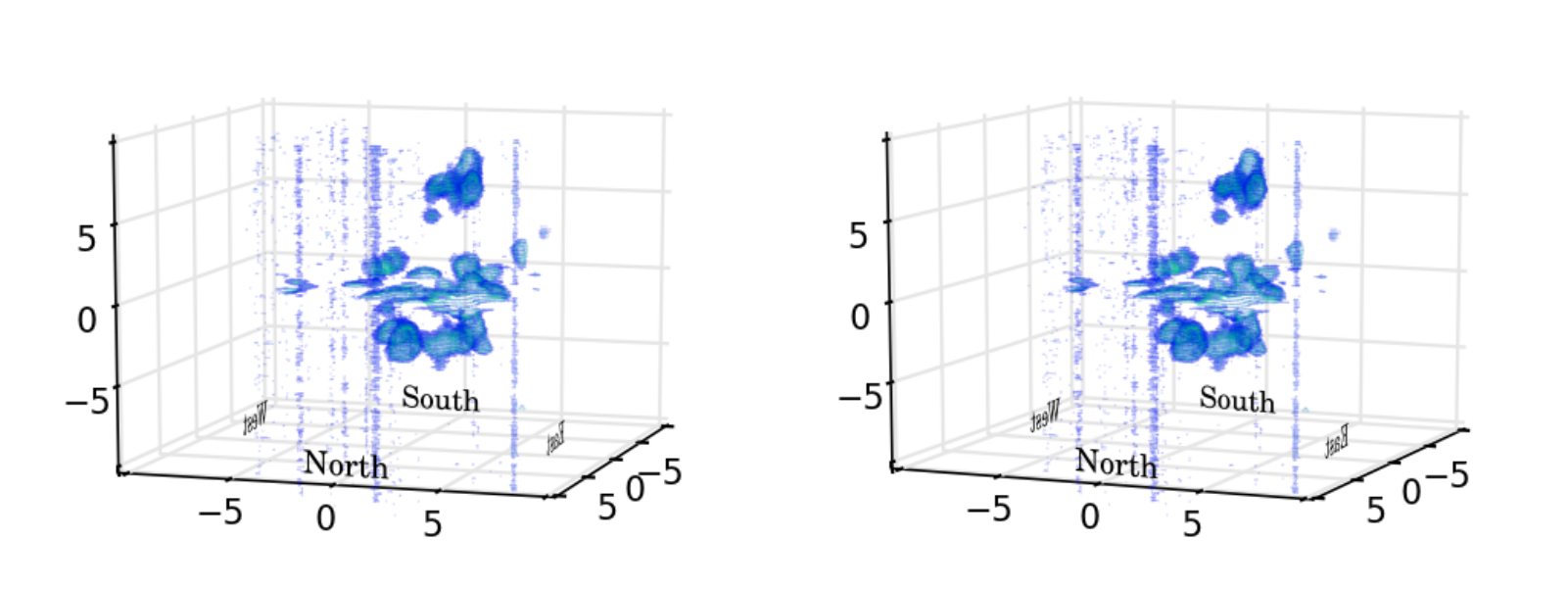}}
\centerline{\includegraphics[scale=0.55]{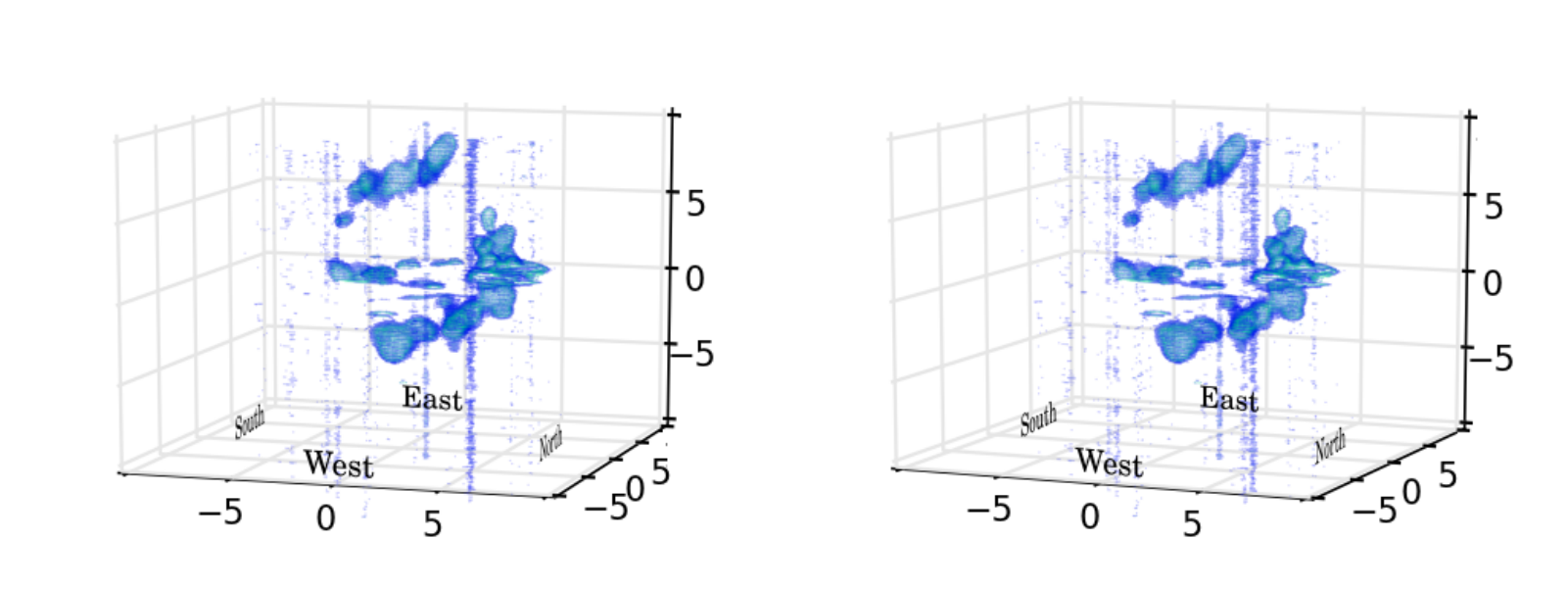}}
\caption{3D stereoscopic maps of the oxygen rich ejecta in SNR N132D, viewed from the NNE and +10$^{\circ}$ elevation with respect to the plane of the sky (top), and from the NNW and +10$^{\circ}$ elevation (bottom). The scales are given in pc. Adapted from \cite{Vogt10c}. }\label{fig:N132D}
\end{figure}

\cite{Vogt10c} used the Wide Field Spectrograph \citep[][]{Dopita07, Dopita10} at Siding Spring Observatory to image the Young Supernova Remnant (YSNR) N132D located in the Large Magellanic Cloud (LMC). The initial data cube axis units are (X [arcsec], Y [arcsec], $\lambda$ [\AA]), and by studying the [O~III] forbidden line at $\lambda$5007 \AA, and its blue- and red-shifted features, they can identify the oxygen-rich knots in the YSNR, and obtain their radial velocities. The data cube axes are then transformed to (X[arcsec], Y[arcsec], v$_\mathrm{r}$ [km s$^{-1}$]). Assuming a distance to the LMC of 50 kpc \citep[from][]{vandenBergh99}, and an age of $\sim$2500 years, they transformed their third data cube axis to a spatial dimension. Thus, they obtained an accurate 3D spatial map with axes (X [pc], Y [pc], Z [pc]) of the oxygen-rich filaments in SNR N132D. Stereo pairs of this 3D map are shown in Fig.~\ref{fig:N132D}.

Stereo pairs are one very useful way to fully understand the true nature of this SNR. \cite{Vogt10c} have used projections of their 3D map, that showed the clumpy structure of the ejecta, as well as hints of the ring-like structure of the ejecta. They also created an interactive 3D map, that enables the reader to zoom, pan, rotate and fly around and through their 3D map. Stereo pairs, together with those others visualization methods, confirmed the presence of the ring structure, and ruled out any perspective effects due to the projection of the 3D map on a 2D plane. The stereo pairs are very efficient in \emph{showing} this ring to the reader, compared to, e.g., a montage of slices. This is a big advantage in the case of SNR N132D, for which the actual shape of the ejecta has been subject to interpretation since the discovery of the remnant \citep[see][]{Lasker80,Morse95,Vogt10c}. Everyone can \emph{see} the ring, and the impression of depth given by stereo pairs is a valuable complement to the interactive 3D map. 

Those stereo pairs have been created with \textsc{Python} using the sTi technique described in Sect.~\ref{Sec:tools}. The whole code, that takes as an input the data cube in a \textsl{.fits} format, contains less than 50 lines, of which only 10 are actually responsible for plotting the data.

\subsection{Theoretical example: relativistic AGN jets and fractal clouds}\label{Sec:Alex}
\begin{figure}[htb!]
\centerline{\includegraphics[scale=0.9]{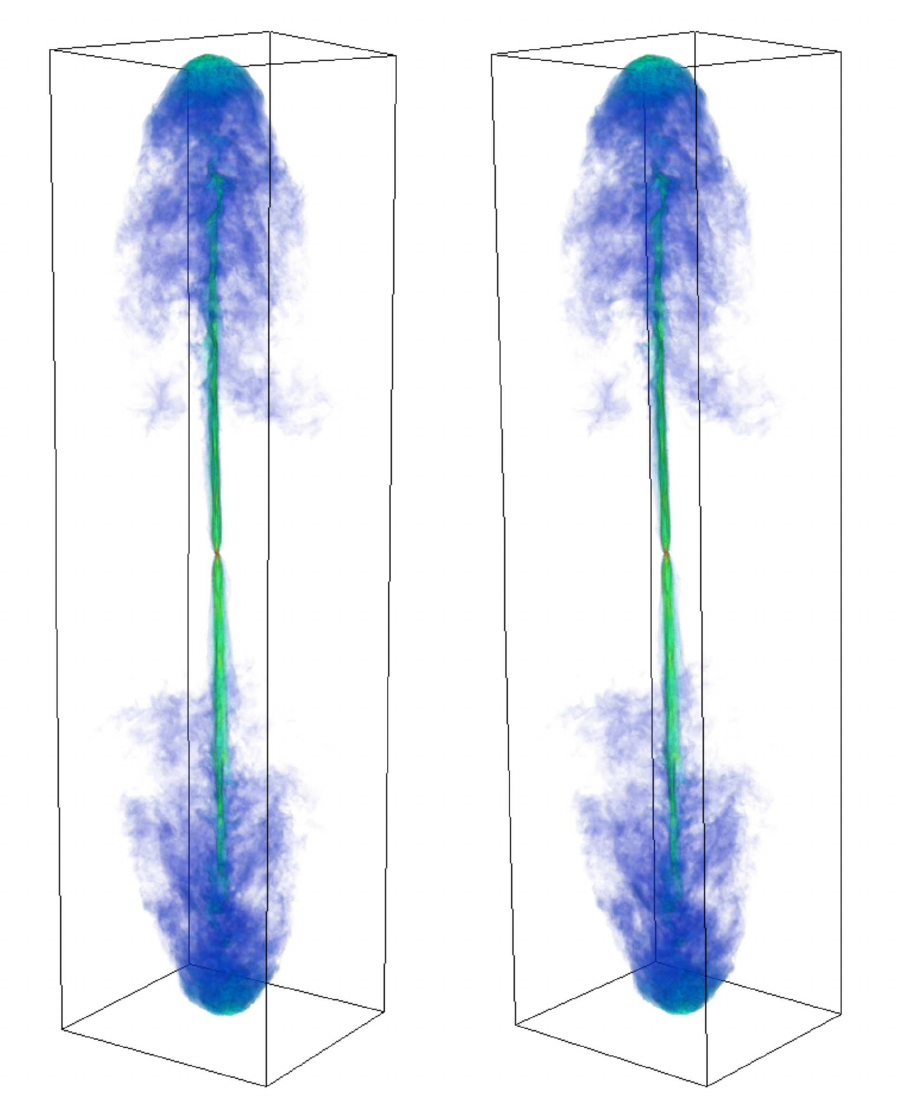}}
\caption{Volume rendered side view of a simulated relativistic AGN jet.}\label{fig:jetlong}
\end{figure}

\begin{figure}[htb!]
\centerline{\includegraphics[scale=0.9]{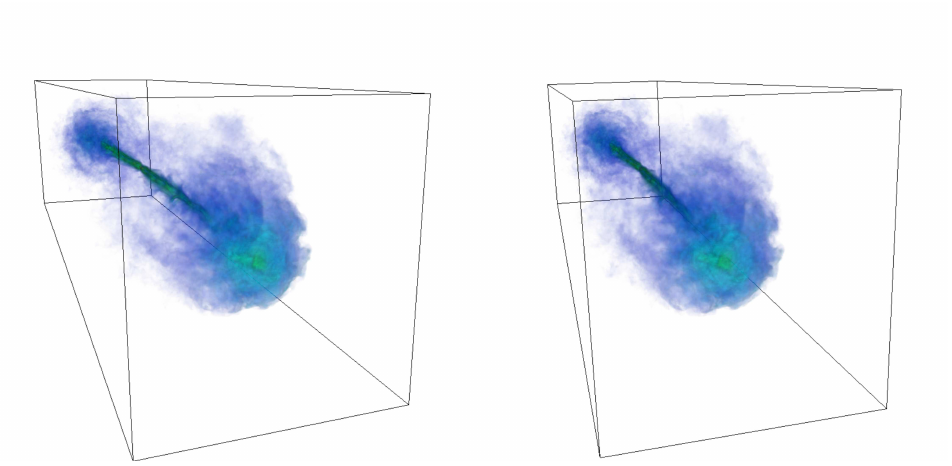}}
\caption{Edge-on view of the same simulated jet shown in Fig.~\ref{fig:jetlong}.}\label{fig:jetshort}
\end{figure}

Stereo pairs can also be used for the visualization of theoretical data sets. In this example, we use stereoscopy to reveal the structure of a simulated relativistic Active Galactic Nucleus (AGN) jet \citep[][]{Wagner10}. The jet simulations were grid-based hydrodynamic simulations which produced multivariate data of thermodynamic quantities, e.g., density, temperature, pressure, velocity components and tracers, as functions of 3 rectilinear spatial coordinates. The resolution of the simulations was $512^3$ cells, each cell representing a physical volume of $(2 \mathrm{pc})^3$. Volume rendered images of the double jet structure are shown in the two stereo pairs, Fig.~\ref{fig:jetlong} and Fig.~\ref{fig:jetshort}. In these renderings, the ray-traced variable was proportional to the 1.8th power of the density and the tracer variable of the jet, which is a measure of the radio emissivity of a jet plasma.


Figure~\ref{fig:jetlong} shows that stereo pairs do not necessarily need to be made up of square images. In this case, the upright rectangular stereo pair allows one to inspect the 3D structure of the jet along the full length of its propagation axis. In particular, one can see the deformations of the central jet stream as it becomes unstable due deceleration and entrainment in the lobes \citep{Bicknell1984}. The structure of the jet lobes along the line of sight is also clearer in the stereo pairs than in either of the 2D images on their own. In the edge-on stereo pairs of Fig.~\ref{fig:jetshort}, the use of stereoscopy enables one to identify the locations of lower and higher concentrations of jet plasma within the volume of the lobe. Globally, one obtains a less ambiguous picture of the true shape and structural characteristics of the complex flow. The viewer obtains a strong sense of depth in the image, despite the contracted view at small angles of the line of sight to the jet axis. The relativistic AGN jet simulation were performed with the \emph{FLASH} code \citep[][]{Fryxell00}. The stereo pairs were produced with the \emph{VisIt} software, developed at the Lawrence Livermore National Laboratory\footnote{VisIt is freely available at: \textsf{https://wci.llnl.gov/codes/visit/home.html}}. \emph{VisIt} has a built-in stereo output.

\begin{figure}[htb!]
\centerline{\includegraphics[scale=1]{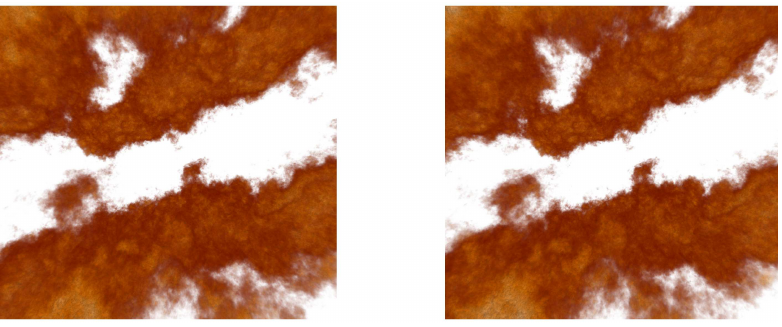}}
\caption{Stereo pair of fractal clouds, viewed from within.}\label{fig:clouds}
\end{figure}

As a further example for the visualization of theoretical data benefiting from stereoscopy, Fig.~\ref{fig:clouds} shows a stereo pair of fractal clouds. These were generated with the procedure outlined by \cite{LA2002}. The procedure creates a spatially fractal distribution in cloud pixels that simultaneously obeys a single-point log-normal probability density function in cloud density. On certain scales, fractal structure and log-normal single point statistics are characteristic of atmospheric \citep{BWP1996} and interstellar \citep{FKS2009} clouds, and 3D data sets such as those depicted in Fig.~\ref{fig:clouds} may be used as initial conditions in hydrodynamical simulations \citep{SBSM2005,SB2007}. Such fractal structures usually prove difficult to visualize. The use of stereo pairs provides an enhanced depth perception and thereby a clearer view of the relative positions of the clouds. The fractal outlines are more obvious, even interior to the clouds.

\section{Future prospects for stereoscopy}\label{Sec:future}

In the previous Sections, we have described how stereoscopy, in the form of stereo pairs, can be a powerful tool for publishing multi-dimensional data sets. As of today, stereoscopy is the \emph{most widely accepted technique for the capture and display of 3D content} \citep[][]{Sharif11}, whereas other methods, such as holography \citep[e.g.][]{Smith75,Ackermann07}, are considered very promising, but, as of now, harder to implement on a large scale. In that sense, we believe that stereo pairs can play a major role in the future of data visualization in Astrophysics (and any field of Science), by providing researchers today with a simple way to explore, discover, imagine, identify, and share new \emph{stereoscopy-based} analysis methods of their data, methods that will then be ready for implementation in exquisite interactive, immersive, high-end visualization tools tomorrow.

Such immersive 3D visualization technologies, both for the scientific and non-scientific community, still appears rather cubersome to use and implement, and often require off-the-shelf, custom software and setup. But technology is moving at a fast pace. For example, working with and sharing data sets in 3D on televisions and hand-held devices might become commonplace fairly soon, as such devices are already easily available on the market. The idea of 3D television is rather old, with early experiments on stereo TV as early as 1920, and the first 3D TV broadcast occurring in 1980 \citep[][]{Onural06}. Yet, very recently, the rapid re-appearance of 3D televisions (and 3D hand-held devices) lead \cite{Sherman10} to state that immersive 3D visualization technologies are now well in the so-called \emph{slope of enlightment} within the Hype's cycle \citep[][]{Fenn08} of new technologies, the last step before reaching a more productive phase. So far, one of the main issues slowing down the expansion of 3D television on the market is most probably the lack of 3D content to be displayed on those devices, a key factor for success \citep[][]{Sharif11}. This is also true for scientific applications of this technology, and a wider usage of stereo pairs may help scientists identify in what ways 3D TV could soon play a significant role in their research. Once the need will have been clearly and widely identified, there is no doubt that the yet missing standardized application programming interface (API) and software links between scientific data sets and already existing hardware will be rapidly implemented. In short, we believe that stereo pairs, an old and well documented tool (which can now be easily implemented), could help scientists keep an open mind, and potentially shape the future of multi-dimensional data visualization and analysis.

\section{Summary}\label{Sec:concl}

We have discussed the concept of stereo pairs and highlighted their potential benefits for the Astrophysics community. First, we presented the free-viewing technique and provided advice to easily visualize both parallel and cross-eyed stereo pairs for the first time. We then argued that stereo pairs can be easily produced and reproduced on a computer screen or on printed material with most of the usual programming languages or software used nowadays within the Astrophysics community. In particular, we have introduced and described an alternative, off-the-shelf, easy way to produce high quality stereo pairs with \textsc{Python}, which we refer to as the \emph{simplified Toe-in} method. This technique adapts the \emph{official} Toe-in procedure taking into account the current limitation of \textsc{Python} plotting abilities, without affecting on the quality of the stereo pairs. Specifically, no vertical parallax can be detected in the resulting stereo pairs. Testing our sTi stereo pairs on several students and astronomers at the Mount Stromlo Observatory revealed that they represent a good trade-off, by being able to convey a satisfactory feeling of depth from any viewpoint, and by being as effective as standard Toe-in stereo pairs with an elevation lower that $|\theta_0|\sim50^{\circ}$. The tests also revealed that sTi stereo pairs provide more depth structure around the data itself as compared to their equivalent Offset stereo pairs, and that the sTi method is in that respect more appropriate for creating stereo pairs in Astrophysics, a fact already observed by \cite{Peterka09}.

We have then used three examples, one idealized and two realistic, with which we have presented various types of stereo pairs, highlighted several aspects of stereoscopic visualization, and identified the main benefits of using stereo pairs as a complement to more standard plotting techniques in a publication. First, they are a polyvalent tool that is adaptable to one's needs; their shape, size, and color can be adapted to best reveal the 3D data set without impacting the ability to transmit a depth perception to the viewer. Second, they profit any multivariate data set, observational or theoretical, and potentially benefit different genres of studies (e.g., of both the theoretical and observational kind). Third, they greatly facilitate the communication of complex 3D shapes. Especially, where a text description might be subject to interpretation, stereo pairs can force upon the viewer a unique view of the data set, thereby avoiding misconceptions. This is possibly the main factor that should dictate the use of stereoscopy in publications and presentations. 

For all these reasons, stereo pairs should be considered a valuable tool for the Astrophysics community - a field where most data sets are multidimensional and multivariate, and where stereo pairs can be applied to many different sub-topics, but always with the common aim of simplifying, clarifying, and eliminating misconceptions. The evolution of informatics has made stereo pairs aesthetic, useful, and straightforward to produce. We are convinced that they have a promising future, given the rapid evolution of 3D visualization hardware and techniques, e.g. 3D televisions. Although we still lack a standardized API and user-friendly software to couple the stereo images to the display devices, these will likely be provided as soon as the need is identified. Sharing astrophysical data sets in 3D on hand-held devices might sound futuristic. Nonetheless, stereo pairs can already be easily stacked into a movie, and played during a talk in a lecture theatre equipped with 3D projection abilities, enabling the audience to experience a glimpse of the 3D future for Astrophysics. In conclusion, we are convinced that the ideas conceived through the ongoing 3D trend currently occurring in the non-scientific community can and ought to be used in Astrophysics. Stereo pairs are a good way to start opening our minds today.

\acknowledgments

We thank the referee for his/her comments that helped greatly improve this paper. This research has made use of NASA's Astrophysics Data System. Part of this research was undertaken on the NCI National Facility at the Australian National University and some software used in this work were in part developed by the DOE-supported ASC / Alliance Center for Astrophysical Thermonuclear Flashes at the University of Chicago.

\appendix

\section{Simplified Toe-in, Toe-in and Offset : a comparison of different stereo pairs creation methods with \textsc{Python}.}\label{App}

We have introduced the simplified Toe-in (sTi) method, described in Sect.~\ref{Sec:tools}, as an alternative to the official Toe-in and Offset methods for creating stereo pairs. The sTi method accounts for the current limitations of \textsc{Matplotlib}, the \textsc{Python} plotting library. The main limitation lies in the fact that the viewpoint towards a 3D plot is defined by only two parameters, the azimuth $\phi$ and the elevation $\theta$. This makes it impossible to create proper Toe-in or Offset stereo pairs directly, for reasons highlighted below (see Appendix~\ref{App:sti}). Nonetheless, our sTi stereo pairs provide an excellent depth perception, especially when $-50^{\circ}\leq\theta\leq50^{\circ}$, and are not more difficult or tiring to visualize as compared to their equivelent Toe-in stereo pairs, according to our small survey of Mount Stromlo students and astronomers. 

In Appendix~\ref{App:comp}, we provide a comparison chart of sTi, Toe-in and Offset stereo pairs for varying elevations. But let us start by describing the issues of sTi stereo pairs, and the required \textsc{Python} code updates to create Toe-in and Offset stereo pairs.

\subsection{sTi stereo pairs}\label{App:sti}

As we defined in Sec.~\ref{Sec:tools}, the LHS and RHS projection viewpoints for the sTi method are located at $S_L(\phi_0-\frac{\delta_0}{2};\theta_0)$ and $S_R(\phi_0+\frac{\delta_0}{2};\theta_0)$, with $2.5^{\circ}\leq\delta_0\leq5^{\circ}$ the opening angle in between the LHS and RHS viewpoint. A schematic of the situation in shown in Fig.~\ref{fig:stiti}, which shows the location of the viewpoints on the visualization sphere. By default in \textsc{Matplotlib}, the sphere is centred on the middle of the data set, and its radius is ten times the size of the data set.

This definition of the sTi LHS and RHS viewpoints makes it easy to use the \textsc{Axes3D} instance, which can take the elevation and azimuth as parameters. However, it introduces two mistakes compared to the official Toe-in method : 
\begin{enumerate}
\item The distance in between the $S_L$ and $S_R$ points, measured along the Great Circle to which they belong (the red line in Fig.~\ref{fig:stiti}), is decreasing with increasing elevation.
\item The orientation of the LHS and RHS views, which by default are oriented towards the z-axis (i.e. along the orange lines) is :
\begin{itemize}
\item not parallel between the LHS and RHS viewpoints.
\item not perpendicular to the red great Circle (as it ought to be).
\item varying with elevation.
\end{itemize}
\end{enumerate}

In other words, with increasing elevation, the sTi LHS and RHS projection points will move along the orange lines, slowly merging towards each other - the cause of the diminishing feeling of depth beyond $\pm50^{\circ}$. The mismatch in the view rotation, increasing from 0 to $\delta_0/2$ at $\theta_0=90^{\circ}$ for each view, is however small enough at any elevation not to be very noticeable. 

\subsection{Toe-in stereo pairs}
To produce Toe-in stereo pairs, we have updated the source code of the \textsc{Axes3d} instance of the \textsc{mplot3d} toolkit, so as to be able to rotate the projection orientation around the view axis. Using spherical trigonometry, one can show that for a central viewpoint $P(\phi_0;\theta_0)$, the LHS and RHS projection need to be made from the position $T_L(\phi_0-\frac{\delta_1}{2};\theta_1)$ and $T_R(\phi_0+\frac{\delta_1}{2};\theta_1)$ with a respective rotation of $\epsilon_T$ and $-\epsilon_T$ around the view axis, where :  
\begin{eqnarray}
\theta_1&=&\frac{\pi}{2}-\arccos\left[\cos(\frac{\delta_0}{2})\cos(\frac{\pi}{2}-\theta_0)\right]\label{eq:1}\\
\delta_1&=&2\times\arccos\left[\left(\cos(\frac{\delta_0}{2})-\cos(\frac{\pi}{2}-\theta_0)\cos(\frac{\pi}{2}-\theta_1)\right)\cdot\frac{1}{\sin(\frac{\pi}{2}-\theta_0)\sin(\frac{\pi}{2}-\theta_1)}\right]\\
\epsilon_T&=&\sin(\frac{\pi}{2}-\theta_0)\cdot\frac{1}{\sin(\frac{\pi}{2}-\theta_1)}\label{eq:3}
\end{eqnarray}

\begin{figure}[htb!]
\centerline{\includegraphics[scale=0.3]{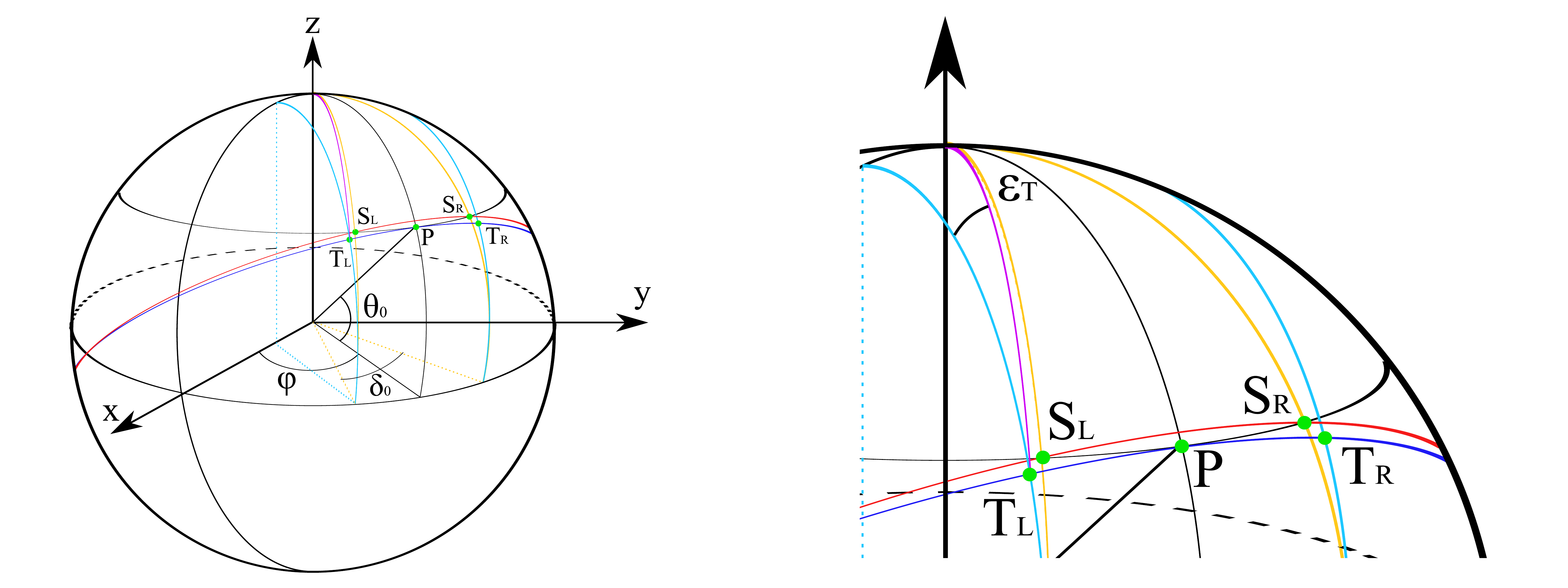}}
\caption{Global (left) and close-up (right) schematics depicting the LHS and RHS camera positions for both an sTi (points $S_L$ and $S_R$) and Toe-in (points $T_L$ and $T_R$) stereo pair. The viewpoints associated with the sTi method (resp. Toe-in method) move along the orange (respect. light blue) line with varying elevation.}\label{fig:stiti}
\end{figure}

The $T_L$ and $T_R$ viewpoints are shown in Fig.~\ref{fig:stiti}. They will move along the light blue lines, always keeping a fixed $\delta_0$ separation at any elevation, as measured along the Great Circle they belong to (dark blue line). The rotation error $\pm\epsilon_T$ of the views (initially oriented along the purple line, but which is corrected by the introduction of a rotation matrix within the plotting source code) increases with elevation, and is as high as 90$^{\circ}$ for $\theta_0=90^{\circ}$. Because the sTi viewpoints converge towards each other at high elevations, their respective projection's rotation error remains small (~$\sim\delta_0/2$) compared to the Toe-in projections - and hence for high elevations we do not require an update of the plotting source code to produce comfortable sTi stereo pairs!

\subsection{Offsets stereo pairs}
Implementing the Offset method is more complicated in \textsc{Python}, and requires much more involved modifications of the plotting code. Instead, we adopted the following, more simplistic, approach. We applied the following transformation to our data cube : 
\begin{eqnarray}
(x;y;z)&\rightarrow&\left(x-\eta\sin(\phi_0);y+\eta\cos(\phi_0);z\right)\\
(x;y;z)&\rightarrow&\left(x+\eta\sin(\phi_0);y-\eta\cos(\phi_0);z\right)
\end{eqnarray}
where  $\eta$ is a scale factor defining the intensity of the offset. In words, we apply a linear translation to the data, perpendicular to the view axis. This method has the disadvantage to \emph{disconnect} the data from the axes, however, it is enough to get an idea of the quality of Offset stereo pairs (and observe that the data itself has less depth content than in Toe-in stereo pairs).

\subsection{Comparison table : sTi vs Toe-in vs Offset stereo pairs}\label{App:comp}

\begin{figure}[htb!]
\centerline{\includegraphics[scale=1]{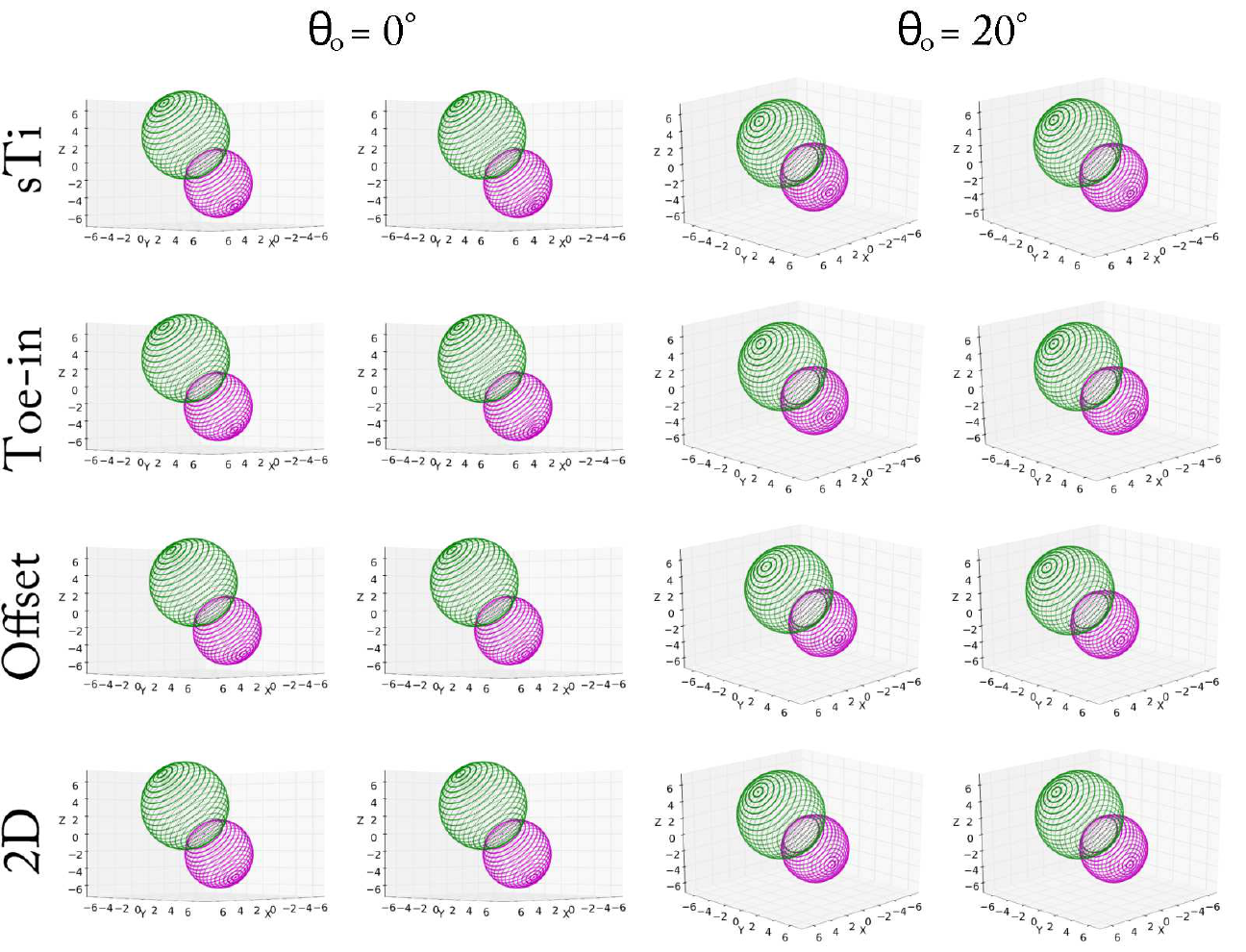}}
\caption{STi, Toe-in, Offset and control stereo pairs for an azimuth $\phi_0=45^{\circ}$ and elevations $\theta_0=0^{\circ}$ and $\theta_0=20^{\circ}$.}\label{fig:comp_1}
\end{figure}

\begin{figure}[htb!]
\centerline{\includegraphics[scale=1]{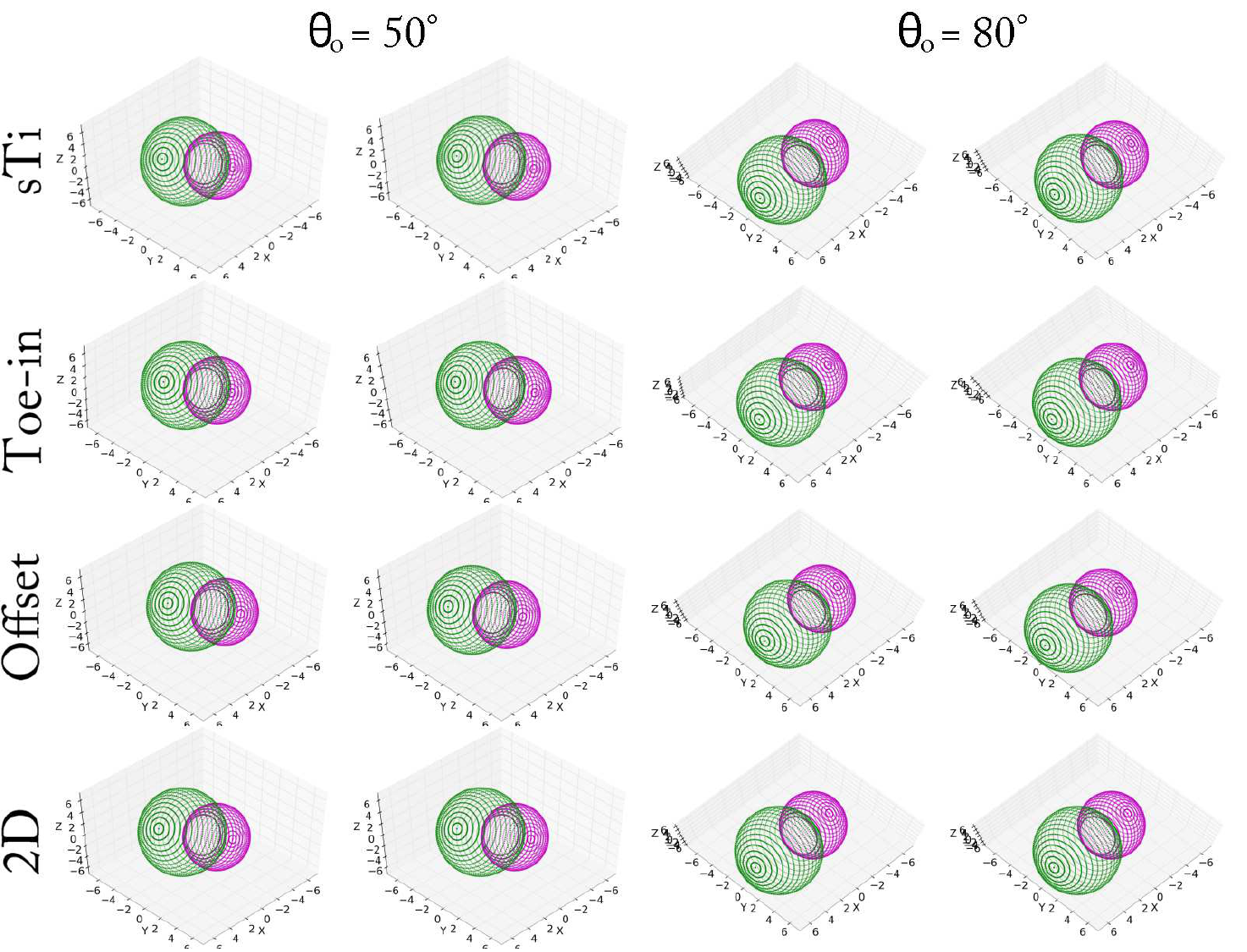}}
\caption{Idem as Fig.~\ref{fig:comp_1}, but for $\theta_0=50^{\circ}$ and $\theta_0=80^{\circ}$}\label{fig:comp_2}
\end{figure}

In Fig.~\ref{fig:comp_1} and Fig.~\ref{fig:comp_2}, we present stereo pairs produced with \textsc{Python} using the sTi, Toe-in, and Offset method described previously. Furthermore, we also include a \emph{control} stereo pair, where we have set identical LHS and RHS images, and that consequently contains no depth information at all. We are aware that by looking long enough at those stereo pairs, the brain will start inducing a \emph{wrong} depth perception, not directly present in the image\footnote{ having the stereo pairs in color might also induce some additional depth perception for this comparative study, but we decided that they might help the non-experienced stereo pairs user to \emph{see} and \emph{feel} our conclusions - the main aim of this Appendix.}; the control stereo pair hence tests that one does not \emph{guess}, rather than \emph{see}, depth information. All stereo pairs have been produced at an azimuth $\phi_0=45^{\circ}$, and varying values for the elevation $\theta_0\in[0^{\circ};20^{\circ};50^{\circ};80^{\circ}]$. In every case, the green (big) sphere is on top, and the symmetry axis of the system lies in the XZ plane.

Comparing the Offset and Toe-in stereo pairs first, one notices that, as expected, the double sphere structure appears to hover over the axes (located further away) in the former ones. This a direct consequence of our simplistic implementation of the Offset method - it is accurate for the data points, but not for the background axes. Nonetheless, focusing on the double-sphere structure itself, it has noticeably less depth elongation in the case of the Offset method. In the Toe-in stereo pairs, not only do the axes wrap around the data points nicely, but the data points themselves appear with a strong feeling of depth. In that sense, we believe the Toe-in method to be much more appropriate for the publication of Astrophysical data cubes, as it provides more depth structure to the data itself.

Comparing the sTi stereo pairs with their equivalent Toe-in pairs, it can be seen that depth perception within the sTi image is gradually degraded as the elevation increases. As mentioned above, this is due to the fact that the LHS and RHS sTi projection viewpoints gradually move towards each other at higher elevation. Nonetheless, some degree of depth perception is present, even as high as $\theta_0=80^{\circ}$. The different orientation of the LHS and RHS images are also staying small, and hence do not need to be corrected by any modification of the \textsc{Matplotlib} source code. In summary, our suggested sTi plotting method, if theoretically not correct, nevertheless appears in practice to provide (very) satisfactory results, and can be directly implemented using \textsc{Python}, \textsc{Matplotlib} and \textsc{mplot3d}. 

The experienced Python user might find it easy to update his source code to create correct Toe-in stereo pairs. We intend to include our code update in future releases of \textsc{Matplotlib}. Until we manage to do so,  we are happy to provide our source code modifications to the interested user directly (which does not require extensive knowledge of \textsc{Python} to be implemented); simply contact F.V. at \emph{fvogt@mso.anu.edu.au}.

\end{document}